\documentclass[prd,nofootinbib,preprint]{revtex4}

\usepackage[english]{babel}
\usepackage{amsmath,amssymb}
\usepackage[dvips]{graphicx}

\newcommand{\eVq}  {\text{eV}^2}

\newcommand{\Dmq}    {\Delta m^2}
\newcommand{\BDmq}   {\Delta\overline{m}^2}
\newcommand{\Bnu}    {\overline{\nu}}
\newcommand{\Bsn}    {\overline{s}}
\newcommand{\Bcs}    {\overline{c}}
\newcommand{\Btheta} {\overline{\theta}}

\newcommand{\Txt}[1] {{\text{#1}}}

\allowdisplaybreaks

\begin{document}

\preprint{YITP-SB-28-03}
\preprint{IFIC/03-27}
\preprint{TUM-HEP-516/03}

\vspace*{2cm}
\title{Status of the CPT Violating Interpretations of the LSND Signal}

\author{M. C. Gonzalez-Garcia}
\email{concha@insti.physics.sunysb.edu}
\affiliation{Y.I.T.P., SUNY at Stony Brook, Stony Brook, NY
  11794-3840, USA,
  and\\
  IFIC, Universitat de Val\`encia - C.S.I.C., Apt 22085, E-46071
  Val\`encia, Spain}
\author{M. Maltoni}
\email{maltoni@ific.uv.es}
\affiliation{IFIC, Universitat de Val\`encia - C.S.I.C., Apt 22085,
  E-46071 Val\`encia, Spain}
\author{T. Schwetz}
\email{schwetz@ph.tum.de}
\affiliation{Institut f{\"u}r Theoretische Physik, Physik Department\\
  Technische Universit{\"a}t M{\"u}nchen,
  James-Franck-Str., D--85748 Garching, Germany
  \vspace*{2cm}
  }

\begin{abstract}
    We study the status of the CPT violating neutrino mass spectrum
    which has been proposed to simultaneously accommodate the
    oscillation data from LSND, KamLAND, atmospheric and solar
    neutrino experiments, as well as the non-observation of
    anti-neutrino disappearance in short-baseline reactor experiments.
    We perform a three-generation analysis of the global data with the
    aim of elucidating the viability of this solution. We find no
    compatibility between the results of the oscillation analysis of
    LSND and all-but-LSND data sets below 3$\sigma$ CL. Furthermore,
    the global data without LSND show no evidence for CPT violation:
    the best fit point of the all-but-LSND analysis occurs very close
    to a CPT conserving scenario.
\end{abstract}

\maketitle

\section{Introduction}

The joint explanation of the oscillation signals observed in LSND~\cite{lsnd},
in solar~\cite{chlorine,sage,gallex,gno,sksollast,snoccnc} and
atmospheric~\cite{skatmlast,skatmpub,macro,soudan} neutrino experiments, and
in the KamLAND reactor experiment~\cite{kamland} provides a big challenge to
neutrino phenomenology.  In Refs.~\cite{cpt1,cpt2} it was observed that the
LSND signal could be accommodated with the solar and atmospheric neutrino
anomalies without enlarging the neutrino sector if CPT was violated. Once such
a drastic modification of standard physics is accepted, oscillations with four
independent $\Dmq$ are possible, two in the neutrino and two in the
anti-neutrino sector. The basic realization behind these proposals is that the
oscillation interpretation of the solar results involves oscillations of
electron neutrinos with $\Dmq_\odot\lesssim 10^{-4}~\eVq$~\cite{solarprekam},
while the LSND signal for short-baseline oscillations with
$\Dmq_\Txt{LSND}\gtrsim 10^{-1}~\eVq$ stems dominantly from anti-neutrinos
($\Bnu_\mu \rightarrow \Bnu_e$). If CPT was violated and neutrino and
anti-neutrino mass spectra and mixing angles were
different~\cite{cpt1,cpt2,cpt4,cpt5,cpt3} both results could be made
compatible in addition to the interpretation of the atmospheric neutrino data
in terms of oscillations of both $\nu_\mu$ and $\Bnu_\mu$ with
$\Dmq_\Txt{atm}\sim 10^{-3}~\eVq$~\cite{skatmlast}.

In the original spectrum proposed, neutrinos had mass splittings
$\Dmq_\odot = \Dmq_{21}\ll\Dmq_{31} = \Dmq_\Txt{atm}$ to explain the
solar and atmospheric observations, while for anti-neutrinos
$\Dmq_\Txt{atm} = \BDmq_{21}\ll\BDmq_{31} = \Dmq_\Txt{LSND}$. Within
this spectrum the mixing angles could be adjusted to obey the relevant
constraints from laboratory experiments, mainly due to the
non-observation of reactor $\Bnu_e$ at short
distances~\cite{chooz,bugey}, and a reasonable description of the data
could be achieved~\cite{cpt3,solveig,strumia}.  In general, stronger
constraints on the possibility of CPT violation arise, once a specific
source of CPT violation which involves other sectors of the theory is
invoked~\cite{cptirina}. For a summary of recent theoretical work and
experimental tests see, for example, Ref.~\cite{kostelecky} and
references therein.

On pure phenomenological grounds, the first test of this scenario came
from the KamLAND~\cite{kamland} experiment since the suggested
CPT-violating neutrino spectrum allowed to reconcile the solar,
atmospheric and LSND anomalies, but, once the constraints from reactor
experiments were imposed, no effect in KamLAND was predicted. The
observation of a deficit in KamLAND at $3.5\sigma$ CL clearly
disfavoured these scenarios.  Furthermore, KamLAND results demonstrate
that $\Bnu_e$ oscillate with parameters consistent with the LMA
$\nu_e$ oscillation solution of the solar anomaly. This fact by itself
can be used to set constraints on the possibility of CPT
violation~\cite{cptjb,strumia,cptmura}.  Within the present KamLAND accuracy,
however, the bounds are not very strong because KamLAND data does not
show a significant evidence of energy distortion.

The present situation is that the results of solar experiments in
$\nu$ oscillations, together with the results from KamLAND and the
bounds from other $\bar\nu$ reactor experiments show that both
neutrinos and anti-neutrinos oscillate with
$\Dmq_\odot,\Dmq_\Txt{reac}\leq 10^{-3}~\eVq$. Adding this to the
evidence of oscillations of both atmospheric neutrinos and
anti-neutrinos with $\Dmq_\Txt{atm} \sim 10^{-3}~\eVq$, leaves no room
for oscillations with $\Dmq_\Txt{LSND}\sim 1~\eVq$. The obvious
conclusion then is that CPT violation can no-longer explain LSND and
perfectly fit all other data~\cite{strumia}.

\begin{figure} \centering
    \includegraphics[width=3.5in]{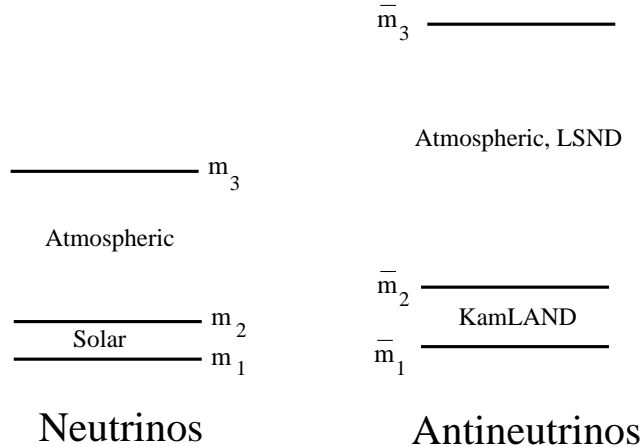}
    \caption{\label{fig:cptschemes}%
      Post KamLAND CPT violating neutrino mass spectrum proposed in
      Ref.~\cite{cpt5}.}
\end{figure}

This conclusion relies strongly on the fact that atmospheric
oscillations have been observed for both neutrinos and anti-neutrinos
with the same $\Dmq_\Txt{atm}$. However, atmospheric neutrino
experiments do not distinguish neutrinos from anti-neutrinos, and
neutrinos contribute more than anti-neutrinos to the event rates by a
factor $\sim$ 4--2 (the factor decreases for higher energies). Based
on this fact, in Ref.~\cite{cpt5} an alternative CPT-violating
spectrum was proposed as shown in
Fig.~\ref{fig:cptschemes}.\footnote{This possibility was also
  discussed in the first version of Ref.~\cite{strumia}.} In this scheme
only atmospheric neutrinos oscillate with $\Dmq_\Txt{atm}$ and give
most of the contribution to the observed zenith angular dependence of
the deficit of $\mu$-like events.  Atmospheric $\Bnu_\mu$ dominantly
oscillate with $\Dmq_\Txt{LSND}$ which leads to an almost constant
(energy and angular independent) suppression of the corresponding
events.  For low $\Bnu_\mu$ energies oscillations with
$\Dmq_\Txt{reac}$ can also be a source of zenith-angular dependence.
The claim in Ref.~\cite{cpt5} was that altogether this suffices to
give a good description of the atmospheric data such that the scheme
in Fig.~\ref{fig:cptschemes} can still be a viable solution to all the
neutrino puzzles. This conclusion was contradicted in
Ref.~\cite{strumia} by an analysis of atmospheric and K2K data.
However, according to the authors in Ref.~\cite{cpt5} an important
point to their conclusion was the consideration of the full 3$\nu$ and
3$\Bnu$ oscillations, while the analysis in Ref.~\cite{strumia} was
made on the basis of a 2$\nu$+2$\Bnu$ approximation.

In this article we determine the status of the CPT violating scenario
in Fig.~\ref{fig:cptschemes} as an explanation to the existing neutrino
anomalies. In order to do this, we perform a three-generation global
analysis of the solar, atmospheric, reactor, and long-baseline (LBL)
data, and compare the allowed parameter regions from this analysis to
the ones required to explain the LSND data.  We find that no
consistency between the parameters determined by the analyses of both
data sets appears below 3$\sigma$ CL.

\section{Notation and Data Inputs}

In what follows we label the states as in Fig.~\ref{fig:cptschemes}
with $\Dmq_{ij} = m^2_i - m^2_j$ and $\BDmq_{ij} = \overline{m}^2_i -
\overline{m}^2_j$.  We denote by $U$ and $\overline{U}$ the
corresponding neutrino and anti-neutrino mixing matrix~\cite{MNS}
which we chose to parametrize as~\cite{PDG}
\begin{equation} \label{eq:matrix}
    U=\left(
    \begin{array}{ccc}
	c_{13} c_{12}
	& s_{12} c_{13}
	& s_{13} \\
	-s_{12} c_{23} - s_{23} s_{13} c_{12}
	& c_{23} c_{12} - s_{23} s_{13} s_{12}
	& s_{23} c_{13} \\
	s_{23} s_{12} - s_{13} c_{23} c_{12}
	& -s_{23} c_{12} - s_{13} s_{12} c_{23}
	& c_{23} c_{13}
    \end{array} \right) \,,
\end{equation}
where $c_{ij} \equiv \cos\theta_{ij}$ and $s_{ij} \equiv
\sin\theta_{ij}$ and with an ``over-bar'' for the corresponding
anti-neutrino mixing. In writing Eq.~\eqref{eq:matrix} we take into
account that for our following description it is correct and
sufficient to set all the CP phases to zero.

In the anti-neutrino sector the reactor
experiments~\cite{chooz,bugey,kamland} provide information on the
$\Bnu_e$ survival probability:
\begin{equation} \begin{split} \label{eq:preac}
    P_{ee}^\Txt{reac} &= 1-\Bcs_{13}^4\sin^22\Btheta_{12}
    \sin^2\left(\frac{\BDmq_{21} L}{4 E} \right) 
    \\
    & \hspace{5mm}
    -\sin^22{\Btheta_{13}} 
    \left[\Bcs_{12}^2\sin^2\left(\frac{\BDmq_{31} L}{4 E}
    \right) + \Bsn_{12}^2\sin^2\left(
    \frac{\BDmq_{32} L}{4 E}\right)\right]
    \\[3mm]
    & \simeq
    \begin{cases}
	1-\sin^22\Btheta_{13}
	\sin^2\left(\dfrac{\BDmq_{31}L}{4E}\right)
	& \text{for}\quad \BDmq_{21} L/E \ll 1
	\\[4mm]
	\Bsn_{13}^4 +\Bcs_{13}^4\left[1-\sin^22\Btheta_{12}
	\sin^2\left(\dfrac{\BDmq_{21} L}{4E}\right)\right]
	&
	\text{for}\quad \BDmq_{31} L/E \gg 1
    \end{cases}
\end{split} \end{equation}
In our analysis we include the results from the
KamLAND~\cite{kamland}, Bugey~\cite{bugey} and CHOOZ~\cite{chooz}
reactor experiments. For KamLAND we include information on the
observed anti-neutrino spectrum which accounts for a total of 13 data
points.  Details of our calculations and statistical treatment of
KamLAND data can be found in Ref.~\cite{ourkland}. For reactor
experiments performed at short-baselines we include the constraints
from Bugey~\cite{bugey} and CHOOZ~\cite{chooz} which are the most
relevant for $\BDmq \gtrsim 0.03~\eVq$ and $0.03~\eVq \gtrsim \BDmq
\gtrsim 10^{-3}~\eVq$, respectively. In our analysis of the CHOOZ data
we include their energy binned data which corresponds to 14 data
points (7-bin positron spectra from both reactors, Table 4 in
Ref.~\cite{chooz}) with one constrained normalization parameter. For
the analysis of Bugey data we use a total of 60 data points given in
Fig.~17 of Ref.~\cite{bugey}, where the ratio of the observed number
of events to the one expected for no oscillations is shown for the
three distances 15 m, 40 m, and 90 m. For technical details of our
Bugey analysis see Ref.~\cite{tomas31}.

In the scheme under consideration the probability associated with the
$\Bnu_\mu \to \Bnu_e$ signal in LSND is given by
\begin{equation} \label{eq:plsnd}
    P_\Txt{LSND}\equiv \sin^22\theta_\Txt{LSND}
    \sin^2 \left(\frac{\Dmq_\Txt{LSND} L}{4 E} \right)=
    \Bsn_{23}^2 \sin^2 2\Btheta_{13}
    \,\sin^2 \left(\frac{\BDmq_{31} L}{4 E} \right) \,,
\end{equation}
where we have neglected terms proportional to $\BDmq_{21}$ which are
irrelevant for LSND distances and energies. In Eq.~\eqref{eq:plsnd} we
have introduced the notation
\begin{equation} \label{eq:defparam}
    \Dmq_\Txt{LSND}=\BDmq_{31}
    \,,\quad
    \sin^22\theta_\Txt{LSND}= \Bsn_{23}^2 \sin^2
    2\Btheta_{13} \,,
\end{equation}
which we will use in the presentation of our results. To include LSND we use
the results of Ref.~\cite{church}, based only on the decay-at-rest
anti-neutrino data sample, which has a high sensitivity to the oscillation
signal. The $\chi^2_\Txt{LSND}$ is derived from a likelihood function obtained
from an event-by-event analysis of the data~\cite{church}.  LSND has also
studied the neutrino channel $\nu_\mu\rightarrow \nu_e$ from decay-in-flight
events. The full 1993-1998 data sample leads to an oscillation probability for
neutrinos of $(0.10 \pm 0.16 \pm 0.04)\%$~\cite{lsnd}, which, although
consistent with the anti-neutrino signal, is also perfectly consistent with
the absence of neutrino oscillations at LSND, as required in the CPT violating
scenario. This fact is the first motivation and successful crucial test for
the explanation of the LSND results by CPT violation. In view of the low
statistical significance of the LSND neutrino signal we do not include it in
the analysis.\footnote{We note, however, that because of a slightly different
experimental configuration the data sample obtained from 1993-1995 had a
higher sensitivity to the neutrino signal. From that data alone a 2.6$\sigma$
signal for $\nu_\mu\rightarrow \nu_e$ oscillations was
obtained~\cite{lsnd_dif} which disfavoured the CPT interpretation.}

For the neutrino sector we use information from solar neutrino
experiments and the K2K~\cite{k2kprl} LBL experiment. For the solar
neutrino analysis we use 80 data points.  We include the two measured
radiochemical rates, from the chlorine~\cite{chlorine} and the
gallium~\cite{sage,gallex,gno} experiments, the 44 zenith-spectral
energy bins of the electron neutrino scattering signal measured by the
SK collaboration~\cite{sksollast}, and the 34 day-night spectral
energy bins measured with the SNO~\cite{snoccnc} detector. We take
account of the BP00~\cite{bp00} predicted fluxes and uncertainties for
all solar neutrino sources except for the $^8$B flux which we treat as
a free parameter. For the relevant cases oscillations with $\Dmq_{31}$
are averaged out for solar neutrinos and the survival probability takes
the form:
\begin{equation} \label{eq:psol}
    P^{3\nu,\rm sol}_{ee}
    = s^4_{13}+ c^4_{13} \, P^{2\nu,\rm sol}_{ee}
    (\Dmq_{21},\theta_{12}) \; ,
\end{equation}
where $P^{2\nu,\rm sol}_{ee} (\Dmq_{21},\theta_{12})$ is the survival
probability for 2$\nu$ mixing obtained with the modified matter
density $N_{e}\rightarrow c^2_{13} N_e$.

The results of the analysis of the solar neutrino
data~\cite{solarprekam} imply that $\Dmq_{21}$ has to be small enough
to be irrelevant for the K2K baseline and energy, and the $\nu_\mu$
survival probability at K2K is
\begin{equation} \label{eq:pk2k}
    P^\Txt{K2K}_{\mu\mu}= 1-4 \left(s^4_{23} s^2_{13} c^2_{13}
    + c^2_{13} s^2_{23} c^2_{23}\right)
    \sin^2\left(\frac{\Dmq_{32} L}{4E}\right) \,.
\end{equation}
In the analysis of K2K we include the data on the normalization and
shape of the spectrum of single-ring $\mu$-like events as a function
of the reconstructed neutrino energy. The total sample corresponds to
29 events~\cite{k2kprl}. We bin the data in five 0.5 GeV bins with $0
< E_\Txt{rec} < 2.5$ plus one bin containing all events above $2.5$
GeV.  The details of the analysis can be found in
Ref.~\cite{ourthree}.

Finally, the analysis of atmospheric neutrino data involves
oscillations of both neutrinos and anti-neutrinos, and, in the
framework of 3$\nu$+3$\Bnu$ mixing, matter effects become relevant. We
solve numerically the evolution equations for neutrinos and
anti-neutrinos in order to obtain the corresponding oscillation
probabilities for both $e$ and $\mu$ flavours. In our calculations, we
use the PREM model of the Earth~\cite{PREM} matter density profile. We
include in our analysis all the contained events from the 1489 SK data
set~\cite{skatmlast}, as well as the upward-going neutrino-induced
muon fluxes from both SK and the MACRO detector~\cite{macro}. This
amounts for a total of 65 data points.  More technical descriptions of
our simulation and statistical analysis can be found in
Ref.~\cite{ouratmos}.

\section{Results}

Our basic approach to test the status of the scheme in
Fig.~\ref{fig:cptschemes} as a possible explanation of the LSND
anomaly together with all other neutrino and anti-neutrino oscillation
data is as follows. First, we perform a global analysis of all the
relevant data, but leaving out LSND data. The goal of this analysis is
to obtain the allowed ranges of parameters $\Dmq_\Txt{LSND}$ and
$\sin^2 2\theta_\Txt{LSND}$ as defined in Eq.~\eqref{eq:defparam} from
this all-but-LSND data set. We then compare these allowed regions to
the corresponding allowed parameter region from LSND, and quantify at
which CL both regions become compatible.

In this approach we start by defining the most general $\chi^2$ for
the all-but-LSND data set:
\begin{multline} \label{eq:chi1}
    \chi^2_\Txt{all-but-LSND}(\Dmq_{21}, \Dmq_{31}, \theta_{12},
    \theta_{23},\theta_{13}|
    \BDmq_{21},\BDmq_{31},
    \Btheta_{12},\Btheta_{23},\Btheta_{13})
    = \chi^2_\Txt{sol}(\Dmq_{21},\theta_{12},\theta_{13})
    \\
    + \chi^2_\Txt{K2K}
    (\Dmq_{31},\theta_{23},\theta_{13})
    + \chi^2_\Txt{Bugey+CHOOZ+KLAND}
    (\BDmq_{21},\BDmq_{31},\Btheta_{12},
    \Btheta_{13})
    \\
    + \chi^2_\Txt{ATM}
    (\Dmq_{21},\Dmq_{31},\theta_{12},\theta_{23},\theta_{13}|
    \BDmq_{21},\BDmq_{31}, \Btheta_{12},
    \Btheta_{23},\Btheta_{13}) \,.
\end{multline}
Notice that in this comparison we have not included the constraints
from the non-observation of $\Bnu_\mu \to \Bnu_e$ transitions at
KARMEN~\cite{karmen}, which, by themselves, disfavour part of the LSND
allowed parameter region. The reason for this omission is that we want
to test the status of the CPT interpretation of the LSND signal using
data independent of the ``tension'' between LSND and KARMEN
results~\cite{church}.

We first focus on the parameters $\Dmq_{21}$ and $\theta_{12}$.  These
parameters are dominantly determined by solar neutrino data, which for
any $\theta_{13}$ prefer values of $\Dmq_{21}$ well below the
sensitivity of atmospheric neutrino data. Therefore, solar data are
mostly important to enforce the ``decoupling'' of the $\Dmq_{21}$
oscillations from the problem. In other words, the atmospheric
neutrino analysis can be made without any loss of generality, in the
standard hierarchical approximation for neutrinos, neglecting the
effect $\Dmq_{21}$ but keeping the generic-3$\nu$ dependence on
$\theta_{13}$. Notice that, unlike in the CPT conserving case, in the
relevant ranges of mass differences, $\theta_{13}$ is not bounded by
any ``terrestrial'' experiment. The dominant source of information on
$\theta_{13}$ is atmospheric data (and less important also solar
data), and for this reason we consistently take into account this
parameter in our analysis.  Thus, after the marginalization over
$\Dmq_{21}$ and $\theta_{12}$ Eq.~\eqref{eq:chi1} takes the form
\begin{multline} \label{eq:chi2}
    \chi^2_\Txt{all-but-LSND}(\Dmq_{31},\theta_{23},\theta_{13}|
    \BDmq_{21},\BDmq_{31},
    \Btheta_{12},\Btheta_{23},\Btheta_{13})
    =\chi^2_\Txt{sol,marg12}(\theta_{13})
    \\
    +\chi^2_\Txt{K2K}
    (\Dmq_{31},\theta_{23},\theta_{13})
    +\chi^2_\Txt{Bugey+CHOOZ+KLAND}
    (\BDmq_{21},\BDmq_{31},\Btheta_{12},
    \Btheta_{13})
    \\
    + \chi^2_\Txt{ATM}
    (\Dmq_{31},\theta_{23},\theta_{13}|
    \BDmq_{21},\BDmq_{31}, \Btheta_{12},
    \Btheta_{23},\Btheta_{13}).
\end{multline}

Let us now discuss the information on $\BDmq_{21}$ from reactor
experiments.  The observation of the $\Bnu_e$ deficit in KamLAND
favours $\BDmq_{21}$ values near the best fit $\BDmq_{21}= 7\times
10^{-5}~\eVq$. For such small values oscillations with $\BDmq_{21}$
have no effect for atmospheric neutrinos. Therefore, we will start by
studying the case $\BDmq_{21} \le 10^{-4}~\eVq$ in
Secs.~\ref{sec:all-but-LSND} and \ref{sec:comparison}. We will relax
this assumption in Sec.~\ref{sec:caseB}, where we investigate also a
possible effect of larger values of $\BDmq_{21}$. Notice also, that
the case of small $\BDmq_{21}$ is continuously connected to the CPT
conserving scenario since it allows for CPT conservation in the
``$12$'' sector. For ${\BDmq_{21}}\leq 10^{-4}~\eVq$, one can easily
marginalize over $\BDmq_{21}$ and $\Btheta_{12}$ and
Eq.~\eqref{eq:chi2} further simplifies to
\begin{multline} \label{eq:chi3}
    \chi^2_\Txt{all-but-LSND}(\Dmq_{31},\theta_{23},\theta_{13}|
    \BDmq_{31},\Btheta_{23},\Btheta_{13})
    =\chi^2_\Txt{sol,marg12}(\theta_{13})
    +\chi^2_\Txt{K2K} (\Dmq_{31},\theta_{23},\theta_{13})
    \\
    +\chi^2_\Txt{Bugey+CHOOZ+KLAND,marg$\overline{12}$}
    (\BDmq_{31},\Btheta_{13})
    + \chi^2_\Txt{ATM}
    (\Dmq_{31},\theta_{23},\theta_{13}|
    \BDmq_{31},\Btheta_{23},\Btheta_{13}).
\end{multline}
Finally, we notice that for any value $\BDmq_{31}\gtrsim 10^{-3}~\eVq$
the results from CHOOZ or, for larger values of $\BDmq_{31}$, from
Bugey, imply a strong limit on $\sin^2 2 \Btheta_{13}$, and in order
to obtain the $\Bnu_e$ disappearance observed in KamLAND
$\Btheta_{13}$ has to be small. Within this bound the results of the
atmospheric neutrino analysis are almost independent of the exact
value of $\Btheta_{13}$ and this parameter can be effectively set to
zero in $ \chi^2_\Txt{ATM}$ without any loss of generality.

For the sake of concreteness we present the quantitative results
corresponding to the normal ordering shown in 
Fig.~\ref{fig:cptschemes} for both neutrinos and anti-neutrinos.  We
have verified that the conclusions hold also for the corresponding inverted
orderings either for neutrinos and anti-neutrinos.  Note that solar and
atmospheric data require $|\Dmq_{21}| \ll |\Dmq_{31}|$, and reactor
and atmospheric data (and furthermore LSND) require $|\BDmq_{21}| \ll
|\BDmq_{31}|$. For such hierarchies, the difference between normal and
inverted schemes arises mainly from Earth matter effects in the
propagation of atmospheric neutrinos and anti-neutrinos, and for the
large values of $|\BDmq_{31}|$ required to explain the LSND signal
Earth matter effects are irrelevant in the anti-neutrino channel.
Within the present experimental accuracy, these effects are not
important enough to lead to significant differences in the results of
the atmospheric neutrino analysis for direct and inverted orderings
(see for instance Ref.~\cite{nohier}). 

\subsection{Analysis of all-but-LSND data}
\label{sec:all-but-LSND}

Using all the data described above except from the LSND experiment we
find the following all-but-LSND best fit point:
\begin{align}
    \Dmq_{31} &= 2.8\times 10^{-3}~\eVq &
    \BDmq_{31} &= 2 \times 10^{-3}~\eVq \nonumber
    \\
    \Dmq_{21} &= 5.8\times 10^{-5}~\eVq &
    \BDmq_{21} &= 7.1\times 10^{-5}~\eVq \nonumber 
    \\
    s^2_{23} &= 0.5 &
    \Bsn^2_{23} &= 0.5 \label{eq:bestfit}
    \\
    s^2_{13} &= 0 &
    \Bsn^2_{13} &= 0.01 \nonumber
    \\
    s^2_{12} &= 0.31 &
    \Bsn^2_{12} &= 0.34 \nonumber
\end{align}
with $\chi^2_\Txt{all-but-LSND,min} = 186.5$ for $238-11 = 227$
dof.\footnote{The 10 neutrino parameters shown in
  Eq.~\eqref{eq:bestfit} plus the free solar $^8$B flux give a total of
  11 fitted parameters.} This can be directly compared to the
corresponding analysis in the CPT conserving scenario:
\begin{align}
    \Dmq_{31} = \BDmq_{31} &= 2.6\times 10^{-3}~\eVq \nonumber 
    \\
    \Dmq_{21} = \BDmq_{21} &= 7.1\times 10^{-5}~\eVq \nonumber
    \\
    s^2_{23} = \Bsn^2_{23} &= 0.5 \label{eq:best_cptcons}
    \\
    s^2_{13} = \Bsn^2_{13} &= 0.009 \nonumber
    \\
    s^2_{12} = \Bsn^2_{12} &= 0.31 \nonumber
\end{align}
with $\chi^2_\Txt{all-but-LSND,min} = 187$ for $238-6 = 232$ dof. We
conclude that, allowing for different mass and mixing parameters for
neutrinos and anti-neutrinos, all-but-LSND data choose a best fit
point very close to CPT conservation and maximal $23$ mixing.

\begin{figure} \centering
    \includegraphics[width=6in]{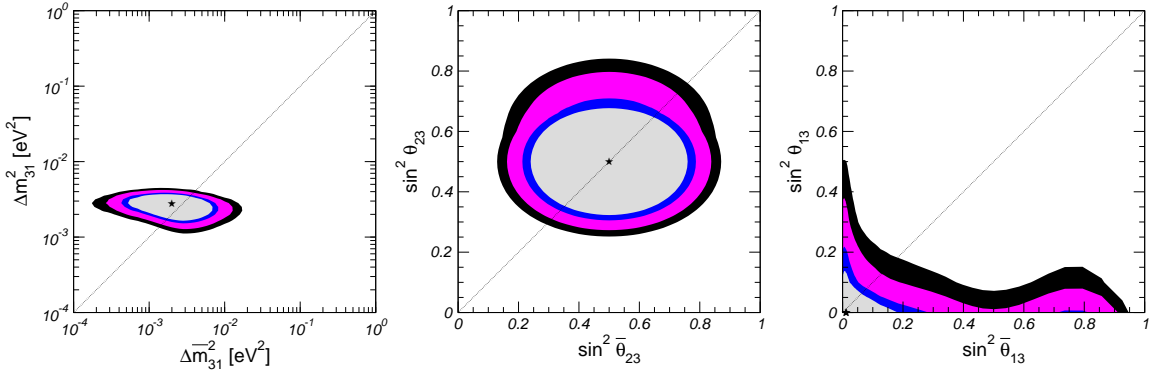}
    \caption{\label{fig:cptlimits}%
      Allowed regions for the largest neutrino and anti-neutrino mass
      splittings $\Dmq_{31}$ and $\BDmq_{31}$ and the mixing angles
      $\theta_{23}$ and $\Btheta_{23}$, and $\theta_{13}$ and $\Btheta_{13}$
      (after marginalization with respect to the undisplayed parameters) for
      $\Dmq_{21}, \BDmq_{21}\leq 10^{-4}~\eVq$ (see text for details).  The
      different contours correspond to the two-dimensional allowed regions at
      90\%, 95\%, 99\% and 3$\sigma$ CL from all-but-LSND data. The best fit
      point is marked with a star.}
\end{figure}

Next we illustrate the amount of CPT violation which is still viable.
In order to do so we plot in Fig.~\ref{fig:cptlimits} the allowed
regions for the largest neutrino and anti-neutrino mass splittings
$\Dmq_{31}$ versus $\BDmq_{31}$ and the mixing angles $\theta_{23}$
versus $\Btheta_{23}$ and $\theta_{13}$ versus $\Btheta_{13}$ and
(after marginalization with respect to all the undisplayed
parameters). The different contours correspond to regions allowed at
90\%, 95\%, 99\% and 3$\sigma$ CL for 2 dof ($\Delta\chi^2 = 4.61$,
$5.99$, $9.21$, $11.83$, respectively).  In general the regions are
larger for anti-neutrino parameters as a consequence of their smaller
contribution to the atmospheric event rates. In particular
$\BDmq_{31}$ can take values below the region of sensitivity of CHOOZ.
As a consequence the limit on $\Btheta_{13}$ at high confidence level
is very week. Our results for the allowed regions for the largest
neutrino and anti-neutrino mass splittings show good qualitative
agreement with the 2$\nu$ analyses of Refs.~\cite{strumia,skatmlast}.
In particular we find that within the 2$\nu$ oscillation approximation
the atmospheric neutrino analysis rejects the CPT violating scenario
at a level close to 4$\sigma$ ($\chi^2 (\BDmq_\Txt{atm} =
\Dmq_\Txt{LSND}) - \chi^2 (\BDmq_\Txt{atm} = \Dmq_\Txt{atm}) = 14$)
which is in reasonable agreement with the $\sim$ 5$\sigma$ rejection
obtained in Ref.~\cite{strumia}. As expected, the introduction of the
3$\nu$ mixing and the reactor data leads to some quantitative
differences in the size of the allowed regions.

From our results shown in Eqs.~\eqref{eq:bestfit},
\eqref{eq:best_cptcons} and in Fig.~\ref{fig:cptlimits} we conclude
that current global neutrino oscillation data excluding LSND show no
evidence for CPT violation, since the best fit point is very close to
a CPT conserving scenario. However, from present data a sizable amount
of CPT violation by neutrino parameters is allowed (for a recent 
discussion on the comparison with the limits existing on the $K - \overline{K}$
mass difference~\cite{PDG} see Ref.~\cite{cptmura}). 
We note that it will be possible to
significantly improve the limits on CPT violation in the neutrino
sector by future experiments such as neutrino factories, see for
example Ref.~\cite{cptnufact}.

\begin{figure} \centering
    \includegraphics[width=3.5in]{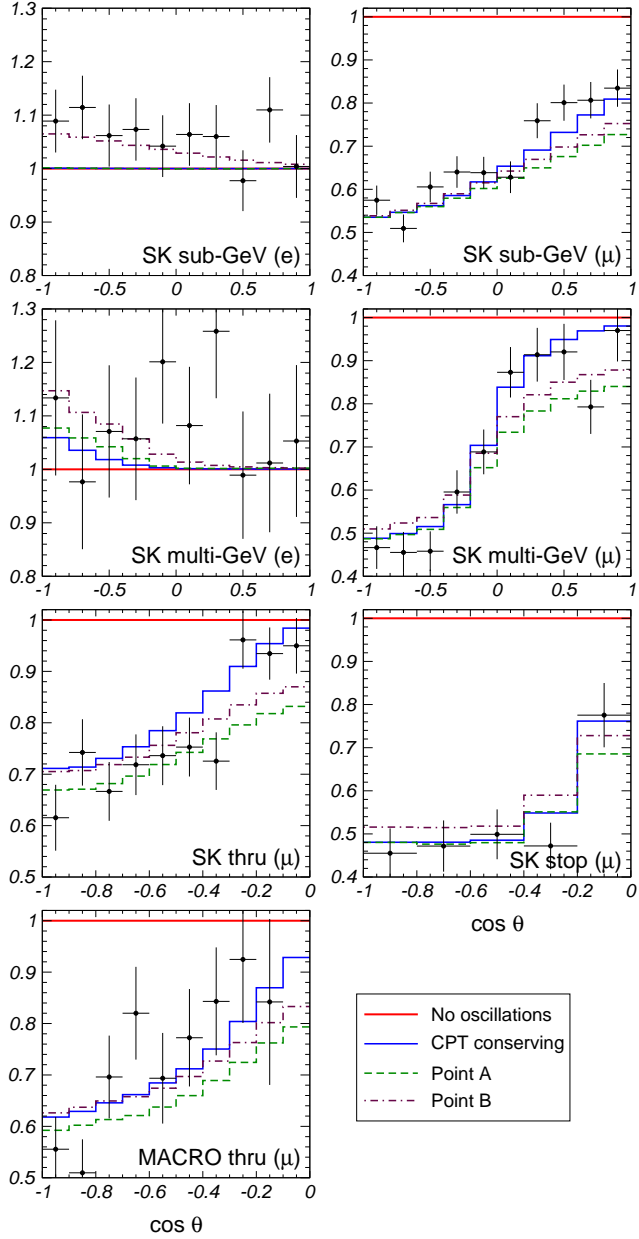}
    \caption{\label{fig:atmdist} %
      Zenith-angle distributions (normalized to the no-oscillation
      prediction) for the Super--Kamiokande $e$-like and $\mu$-like
      contained events, for the Super--Kamiokande stopping and
      through-going muon events and for Macro up-going muons.  The full
      line gives the distribution for the best fit of
      $\nu_\mu\rightarrow\nu_\tau$ oscillations and CPT conservation:
      $\Dmq_{31} = \BDmq_{31} = 2.6\times 10^{-3}~\eVq$, $s^2_{23} =
      \Bsn_{23}^2 = 0.5$, $s^2_{13} = \Bsn_{13}^2 = 0$, and $\Dmq_{21}
      = \BDmq_{21} \lesssim 10^{-4}~\eVq$. The lines label as ``Point
      A'' and ``Point B'' are the expected distributions for typical
      LSND--compatible CPT violating cases with the following
      parameter values: Point A: $\Dmq_{31} = 2.5\times 10^{-3}~\eVq$,
      $\BDmq_{31} = 0.9~\eVq$, $s^2_{23} = \Bsn_{23}^2=0.5$,
      $s^2_{13}=0.05$, $\Bsn_{13}^2 = 0.005$, $\Dmq_{21} \lesssim
      10^{-4}~\eVq$, and $\BDmq_{21} \lesssim 10^{-4}~\eVq$; Point B:
      $\Dmq_{31} = 2.5\times 10^{-3}~\eVq$, $\BDmq_{31} = {\cal
      O}(\eVq)$, $s^2_{23} = 0.5$, $\Bsn_{23}^2 = 0.25$, $s^2_{13} =
      0.05$, $\Bsn_{13}^2 = 0.005$, $\Dmq_{21} \lesssim 10^{-4}~\eVq$,
      $\BDmq_{21} = 5\times 10^{-4}~\eVq$, and $\Bsn_{12}^2 = 0.75$.}
\end{figure}

Concerning LSND, we find that values of $\BDmq_{31} =
\BDmq_\Txt{LSND}$ large enough to fit the LSND result do not appear as
part of the 3$\sigma$ CL allowed region of the all-but-LSND analysis
which is bounded to $\BDmq_{31} < 1.6 \times 10^{-2}~\eVq$.  The upper
bound on $\BDmq_{31}$ is determined by atmospheric neutrino data (and
slightly strengthened by the reactor constraints). To illustrate the
physics behind this result we show in Fig.~\ref{fig:atmdist} the
zenith-angle distributions of various atmospheric data samples for
``Point A'' with the following parameter values: $\Dmq_{31} =
2.5\times 10^{-3}~\eVq$, $\BDmq_{31} = 0.9~\eVq$, $s^2_{23} =
\Bsn_{23}^2 = 0.5$, $s^2_{13} = 0.05$, $\Bsn_{13}^2 = 0.005$,
$\Dmq_{21}\lesssim 10^{-4}~\eVq$, and $\BDmq_{21} \lesssim
10^{-4}~\eVq$. This point has been chosen to be compatible with the
LSND result while keeping an optimized $\chi^2_\Txt{all-but-LSND}$. As
seen in the figure this point fails in reproducing the up-down
asymmetry of multi-GeV muons as a consequence of the
angular-independence in the deficit of the anti-neutrino events.
Furthermore, it predicts a too large deficit of up-going muon events
near the horizon since $\Bnu_\mu$ oscillations with $\BDmq_{31} =
0.9~\eVq$ lead to the disappearance of $\Bnu_\mu$'s even at those
higher energies and shorter distances. For up-going muons the
contribution from anti-neutrino events is only half of that from
neutrino events. As a consequence this data sample is most sensitive
to the anti-neutrino oscillation parameters. Both effects, the wash
out of the up-down asymmetry and the deficit of horizontally arriving
up-going muons, contribute in comparable amounts to the statistical
disfavouring of the CPT violating scenario.

\subsection{Comparison of the all-but-LSND and the LSND data sets}
\label{sec:comparison}

\begin{figure} \centering
    \includegraphics[width=4in]{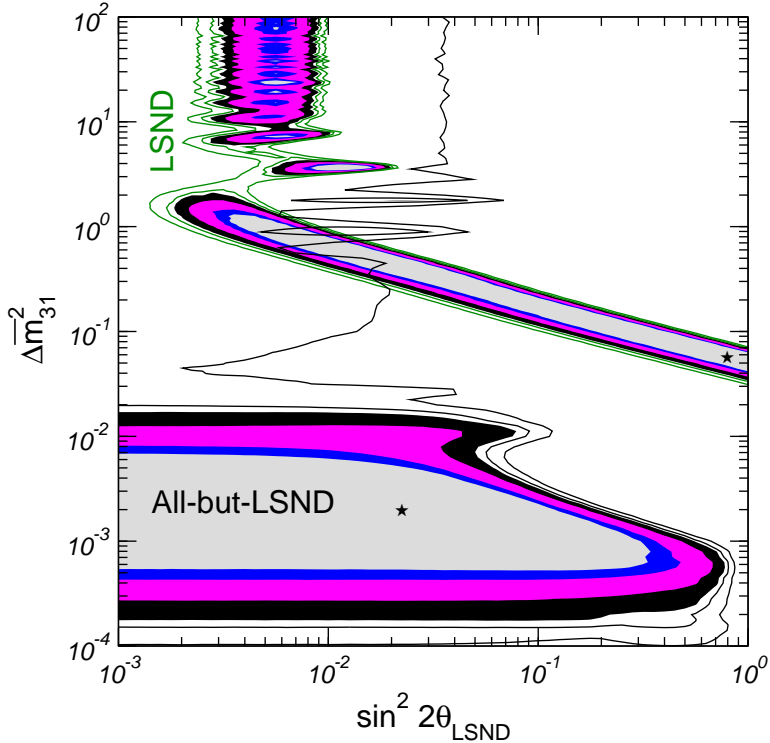}
    \caption{\label{fig:cptstatus}%
      90\%, 95\%, 99\%, and $3\sigma$ CL allowed regions (filled) in
      the ($\BDmq_{31} = \Dmq_\Txt{LSND}$, $\sin^2
      2\theta_\Txt{LSND}$) plane required to explain the LSND signal
      together with the corresponding allowed regions from our global
      analysis of all-but-LSND data. The contour lines correspond to
      $\Delta\chi^2 = 13$ and 16 (3.2$\sigma$ and 3.6$\sigma$,
      respectively).}
\end{figure}

It is clear from these results that the CPT violation scenario cannot
give a good description of the LSND data and simultaneously fit
all-but-LSND results. The quantification of this statement is
displayed in Fig.~\ref{fig:cptstatus} where we show the allowed
regions in the ($\BDmq_{31} = \Dmq_\Txt{LSND}$, $\sin^2
2\theta_\Txt{LSND}$) plane required to explain the LSND signal
together with the corresponding allowed regions from our global
analysis of all-but-LSND data.

Fig.~\ref{fig:cptstatus} illustrates that below 3$\sigma$ CL there is
no overlap between the allowed region of the LSND analysis and the
all-but-LSND one, and that for this last one the region is restricted
to $\BDmq_{31} = \Dmq_\Txt{LSND}<0.02~\eVq$.  At higher CL values of
$\BDmq_{31}\sim {\cal O}(\eVq)$ become allowed -- as determined mainly
by the constraints from Bugey -- and an agreement becomes possible. We
find that in the neighbourhood of $\BDmq_{31} = \Dmq_\Txt{LSND} =
0.9~\eVq$ and $\sin^2 2\theta_\Txt{LSND}= 0.01$ the LSND and the
all-but-LSND allowed regions start having some marginal agreement
slightly above 3$\sigma$ CL (at $\Delta\chi^2 = 12.2$).  A less
fine-tuned agreement appears at 3.3$\sigma$ CL ($\Delta\chi^2\sim
14)$ for $\BDmq_{31} = \Dmq_\Txt{LSND}\gtrsim 0.5~\eVq$ and $\sin^2
2\theta_\Txt{LSND}\lesssim 0.01$ .

Alternatively the quality of the joint description of LSND and all the
other data can be evaluated by performing a global fit based on the
total $\chi^2$-function $\chi^2_\Txt{tot} = \chi^2_\Txt{all-but-LSND}
+ \chi^2_\Txt{LSND}$, and applying a goodness-of-fit test. The best
fit point of the global analysis is $\sin^22\theta_\Txt{LSND} = 6.3
\times 10^{-3}$ and $\BDmq_{31} = 0.89~\eVq$ with
$\chi^2_\Txt{tot,min} = 200.9$. In the following we will use the
so-called parameter goodness-of-fit~\cite{fourlast,PG}, which is
particularly suitable to test the compatibility of independent data
sets.  Applying this method to the present case we consider the
statistic
\begin{equation} \begin{split} \label{eq:chi2PG}
    \bar\chi^2 &\equiv \chi^2_\Txt{tot,min}
    - \chi^2_\Txt{all-but-LSND,min}
    - \chi^2_\Txt{LSND,min}
    \\
    &= \Delta\chi^2_\Txt{all-but-LSND}(\mathrm{b.f.}) +
    \Delta\chi^2_\Txt{LSND}(\mathrm{b.f.}) \,,
\end{split} \end{equation}
where b.f.\ denotes the global best fit point.  The $\bar\chi^2$ of
Eq.~\eqref{eq:chi2PG} has to be evaluated for 2 dof, corresponding to
the two parameters $\sin^22\theta_\Txt{LSND}$ and $\BDmq_{31}=
\Dmq_\Txt{LSND}$ coupling the two data sets: all-but-LSND and LSND
(see Ref.~\cite{PG} for details about the parameter goodness-of-fit).
From $\Delta\chi^2_\Txt{all-but-LSND} = 12.7$ and
$\Delta\chi^2_\Txt{LSND} = 1.7$ we obtain $\bar\chi^2 = 14.4$ leading
to the marginal parameter goodness-of-fit of $7.5\times 10^{-4}$.

\subsection{The effect of large $\BDmq_{21}$}
\label{sec:caseB}

\begin{figure} \centering
    \includegraphics[width=3in]{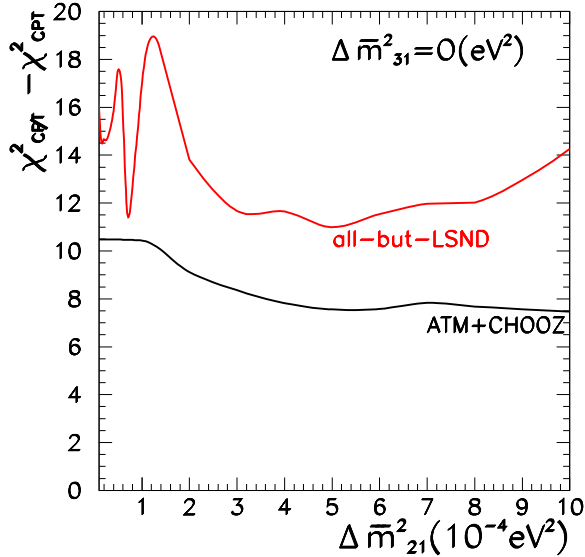}
    \caption{\label{fig:chi2cpt} %
      $\Delta\chi^2 = \chi^2_\Txt{min,C\slash\!\!PT}-\chi^2_\Txt{min,
	CPT}$ as a function of $\BDmq_{21}$ from the
      analysis of atmospheric+CHOOZ data (lower line) and from
      all-but-LSND (upper line) data (see text for details). }
\end{figure}

Finally, we study whether the conclusions of the previous subsections
could be affected by allowing for larger values of $\BDmq_{21}$, such
that its effect can show up in the atmospheric neutrino data and
improve the quality of the fit as suggested in Ref.~\cite{cpt5}. In
Fig.~\ref{fig:chi2cpt} we show the dependence on $\BDmq_{21}$ of the
$\chi^2$ obtained for the analysis of atmospheric and CHOOZ data, and
for all-but-LSND data.  In each curve we have marginalized with
respect to the undisplayed variables subject to the condition
$\BDmq_{31}\gtrsim 1~\eVq$.  For the sake of normalization we have
subtracted in each case the corresponding $\chi^2_\Txt{min,CPT}$ for
the CPT conserving scenario.

The figure shows that, indeed, considering only atmospheric+CHOOZ data
there is an improvement (although mild) in the quality of the fit due
to the effect of oscillations with larger values of $\BDmq_{21}$. To
illustrate this result we show in Fig.~\ref{fig:atmdist} the
zenith-angle distributions of various atmospheric data samples for
``Point B'', which gives the lowest $\chi^2_\Txt{all-but-LSND}$ for a
larger value of $\BDmq_{21}$: $\Dmq_{31} = 2.5\times 10^{-3}~\eVq$,
$\BDmq_{31} = {\cal O}(\eVq)$, $s^2_{23} = 0.5$, $\Bsn_{23}^2 = 0.25$,
$s^2_{13} = 0.05$, $\Bsn_{13}^2 = 0.005$, $\Dmq_{21}\lesssim
10^{-4}~\eVq$, $\BDmq_{21} = 5\times 10^{-4}~\eVq$, and $\Bsn_{12}^2 =
0.75$. The figure shows that the main effect of $\BDmq_{21}$
oscillations is to increase the number of contained $e$-like events,
in particular sub-GeV~\cite{nohier,smirnov}, improving the fit for
those events. However, the two main sources of discrepancy in the
atmospheric fit in these scenarios -- the small up-down asymmetry for
multi-GeV muon-like events and the deficit of horizontally arriving
up-going muons -- remain a problem even when atmospheric anti-neutrino
oscillations with the two relevant wavelengths are included. We conclude
from this analysis that the claim in Ref.~\cite{cpt5} of a possible
improvement of the atmospheric neutrino fit due to the inclusion of
the effect of oscillations with larger values of $\BDmq_{21}$ is 
qualitatively correct for the contained events, although
quantitatively relevant only for the sub-GeV $e$-like events. 
Moreover, we find that quantitatively the improvement in the fit is not 
enough to make the scenario viable. This conclusion is partially based on 
the bad description of the upward-going muon events in the CPT
violating scenario, a fact which was overlooked  in Ref.~\cite{cpt5}. 

In other words, our results show that atmospheric neutrino data are
precise enough to be sensitive to the anti-neutrino oscillation
parameters, and it cannot be well described by a combination of
neutrino oscillations with $\Dmq_{31}\simeq 3 \times 10^{-3}~\eVq$ and
anti-neutrino two-wavelength oscillations with $\BDmq_{31}\sim {\cal
  O}(\eVq)$ and $\BDmq_{21}\sim$ few $10^{-4}~\eVq$.

The net effect in the global all-but-LSND analysis is that the
improvement in the atmospheric fit is not enough to make the scenarios
viable because it does not fully overcome the preference of smaller
$\BDmq_{21}$ in KamLAND (even within their present limited statistics)
as illustrated in the all-but-LSND curve in
Fig.~\ref{fig:chi2cpt}. The local minimum at $\BDmq_{21} = 7.1\times
10^{-5}~\eVq$ corresponds to a point in the vicinity of the point
where the LSND and all-but-LSND regions in Fig.~\ref{fig:cptstatus}
first meet. From the curve in Fig.~\ref{fig:chi2cpt} we see that the
improvement obtained by moving to the minimum at $\BDmq_{21} = 5\times
10^{-4}~\eVq$ is only 0.5 units in $\chi^2$. We conclude that higher
$\BDmq_{21}$ values do not significantly affect the overall status of
the CPT violating scenario.

\section{Conclusions}

We have explored the possibility of explaining all the existing
neutrino anomalies without enlarging the neutrino sector but allowing
for CPT violation as described by the scenario in
Fig.~\ref{fig:cptschemes}. In order to do so we have performed a
compatibility test between the results of the oscillation analysis of
the LSND on one side and all-but-LSND data on the other in the
framework of 3$\nu$+3$\Bnu$ oscillations.  Our main results are shown
in Fig.~\ref{fig:cptstatus}. We find that the allowed regions for both
data sets have no overlap at 3$\sigma$ CL. Alternatively, using the
so-called parameter goodness-of-fit our results imply that the
probability for compatibility between both data sets within this
scenario is only $7.5\times 10^{-4}$.

The information most relevant to our conclusion comes from the
atmospheric neutrino events. Our results show that, within the
constraints imposed by solar and LBL neutrino data, and reactor
anti-neutrino experiments, atmospheric data are precise enough to be
sensitive to anti-neutrino oscillation parameters and cannot be
described with oscillations with the wavelengths required in the CPT
violating scenario.

Furthermore, the global oscillation data without LSND show no evidence
for any CPT violation. An analysis of the all-but-LSND data set
allowing for different mass and mixing parameters of neutrinos and
anti-neutrinos gives a best fit point very close to perfect CPT
conservation.

\acknowledgments

This work was supported in part by the National Science Foundation grant
PHY0098527. M.C.G.-G.\ is also supported by Spanish Grants No
FPA-2001-3031 and CTIDIB/2002/24. M.M.\ is supported by the Spanish grant
BFM2002-00345, by the European Commission RTN network HPRN-CT-2000-00148,
by the European Science Foundation Neutrino Astrophysics Network No 86,
and by the Marie Curie contract HPMF-CT-2000-01008. T.S.\ is supported by
the ``Sonderforschungsbereich 375-95 f{\"u}r Astro-Teilchenphysik'' der
Deutschen Forschungsgemeinschaft.

\end{document}